\documentclass[letterpaper]{article}
\usepackage{aaai}
\usepackage{times}
\usepackage{helvet}
\usepackage{courier}
\usepackage[ruled]{algorithm2e}

\SetAlFnt{\small}
\SetAlCapFnt{\small}
\SetAlCapNameFnt{\small}
\SetAlCapHSkip{0pt}
\IncMargin{-\parindent}

\usepackage{amsmath}
\usepackage{amssymb}
\usepackage{algorithmicx}
\usepackage{subfigure}
\usepackage{algpseudocode}
\usepackage[hidelinks]{hyperref}
\DeclareMathOperator*{\argmax}{arg\,max}
\usepackage{url}
\usepackage{color}
\usepackage{graphicx}
\begin{document}
\title{From Tweets to Events: Exploring a Scalable Solution for Twitter Streams}
\author{Shamanth Kumar$^\dagger$, Huan Liu$^\dagger$, Sameep Mehta$^\star$, and L. Venkata Subramaniam$^\star$\\
$^\dagger$ Computer Science \& Engineering, CIDSE, Arizona State University, Tempe, AZ\\
$^\star$ IBM India Research Lab, New Delhi, India\\
\{shamanth.kumar, huan.liu\}@asu.edu, \{sameepmehta, lvsubram\}@in.ibm.com\\
}
\maketitle
\begin{abstract}
The unprecedented use of social media through smartphones and other web-enabled mobile devices has enabled the rapid adoption of platforms like Twitter. Event detection has found many applications on the web, including breaking news identification and summarization. The recent increase in the usage of Twitter during crises has attracted researchers to focus on detecting events in tweets. However, current solutions have focused on static Twitter data. The necessity to detect events in a streaming environment during fast paced events such as a crisis presents new opportunities and challenges. In this paper, we investigate event detection in the context of real-time Twitter streams as observed in real-world crises. We highlight the key challenges in this problem: the informal nature of text, and the high volume and high velocity characteristics of Twitter streams. We present a novel approach to address these challenges using single-pass clustering and the compression distance to efficiently detect events in Twitter streams. Through experiments on large Twitter datasets, we demonstrate that the proposed framework is able to detect events in near real-time and can scale to large and noisy Twitter streams.
\end{abstract}


\maketitle
\section{Introduction}
\label{sec:intro}
Social networking sites like Twitter have proven to be popular outlets for information dissemination during crises. It has been observed that information related to a crisis is released on social media sites before traditional news sites~\cite{Gilg-Lee13}, \cite{Kham13}.  During the Arab Spring movement, Twitter was used as an information source to coordinate protests and to bring awareness to the atrocities~\cite{Huan11}. In recent world events, social media data has been shown to be effective in detecting earthquakes~\cite{Saka-etal10}, rumors~\cite{Mend-etal10}, and identifying characteristics of information propagation~\cite{Qu-etal11}. This motivates us to study the problem of event detection, which is an interesting and important problem in this domain.

Event detection approaches designed for documents cannot be directly applied to tweets due to the difference in the characteristics of tweets. Unlike a traditional document stream, a Twitter stream suffers from the informality of language, and differs in both volume and velocity characteristics. Existing approaches to event detection in tweets focus on the problem in an offline setting, where the corpus is static and multiple passes can be employed in the solution. However, event detection in streaming environment presents unique challenges, which prevent the direct application of existing approaches. Detecting events in streaming Twitter data has the following new challenges: \\{\bf Informal use of language:} Twitter users produce and consume information in a very informal manner compared with traditional media~\cite{Pari-etal12}. Mis-spellings, abbreviations, contractions, and slang are rampant in tweets, which is exacerbated by the length restriction (a tweet can have no more than 140 characters).\\
{\bf Noise:} While traditional event detection approaches assume that all documents are relevant, Twitter data typically contains a vast amount of noise and not every tweet is related to an event~\cite{Pear09}.\\
{\bf Dynamicity:} Twitter streams are highly dynamic with high volume and high velocity as typical characteristics. Approximately 400 million tweets are now posted on Twitter every day~\cite{Tsuk13}. Event detection methods need to be scalable to handle this high volume of tweets. 

Social media such as Twitter empower their users to publish information as soon as a real-world event occurs. However, this information is not curated as in the case of traditional documents, such as news article. Whereas, each news article is part of an event, not every tweet is expected to be part of an event, as there is a significant amount of noise and inter-personal communication. In this paper, we address the above challenges through a novel approach which can:
\begin{enumerate}
\item Effectively handle the informality of language in a Twitter stream through the selection of an appropriate distance measure;
\item Efficiently detect events without the need for a preset number of events; and
\item Scale to high volume streaming Twitter data.
\end{enumerate}


\section{Problem Definition}
Given an ordered stream of tweets $T = {t_1,t_2,t_3,...}$, where each $t_i$ is associated with a timestamp indicating its publication time, we need to detect events $E = {e_1, e_2, ...e_m}$, where $e_j = {t_1, t_2, ..., t_k}$, where $t_{k} \in T$ and $j \in [1,m]$.
\textbf{Event:} An event is formally defined as a set of similar tweets $E = {t_1,t_2,...,t_k}$ with high user diversity.\\
\textbf{User Diversity}: of an event is the diversity of the user population who contribute to the event. The intuition here is that a diverse user population lends credibility to the event and helps us filter out noise. Entropy is a measure commonly used to compute the amount of information in a text and here we reformulate it to measure user diversity of an event. Given an event $e$, its \textit{User Diversity} $H(e)$ is
\begin{equation}
H(e) = -\sum\limits_{i}\frac{n_{u_i}}{n}log\frac{n_{u_i}}{n},
\label{eq:interestingness}
\end{equation}
where $u_i$ is the $i$th user in the cluster. Here, $n_{u_i}$ is the number of tweets published by user $u_{i}$, which are part of the cluster $C$, and $n$ is the total number of tweets in the cluster. 
\subsection{Hardness of Event Detection}
To detect events $E$, we aim to find a clustering $C$ of the tweets $T$ such that the user diversity of the resulting clusters and the distance between tweets in different clusters is maximized. Let us assume, that the number of events $m$ is known. Then, we can define the objective function as
\begin{eqnarray}
\argmax \sum_{\substack{e \in E}} (\sum_j D(e,e_j))+H(e)),  ~~~~ s.t. ~|E| = m
\label{eq:obj}
\end{eqnarray}
where $D(e_i,e_j)$ is the maximum distance between any pair of elements in clusters $e_i$ and $e_j$ computed as
\begin{equation}
D(e_i,e_j) = max_{a,b}(D(e_i^{(a)},e_j^{(b)})), ~~~~s.t.~|e_i| = a, |e_j|=b.
\end{equation}

To show that this function is hard, we prove that it is submodular using the following properties:\\
\textbf{Monotone} The function is monotone as at each step a tweet is added to the nearest cluster, hence the summation of the distances between clusters cannot decrease.\\
\textbf{Diminishing Returns:} Let us assume that $S$ and $T$ are two clusters, where $S \subseteq T$. If a tweet $x$ is added to $S$ and $T$, then the change in $D(S,P)$ and $D(T,P)$, where $P$ is any other cluster follows one of two scenarios:
\begin{itemize}
\item If both T and S contain the tweet which maximizes D(.) and $x$ increases the distance to $P$. Then $D(T \cup {x},P) - D(T,P)) = D(S \cup {x},P) = D(S,P)$,
\item Otherwise, if there is a tweet $x' \not\in S$ but $x' \in T$, such that $D(S,P) < D(T,P)$, then $D(T \cup {x},P) - D(T,P) \leq D(S \cup {x},P) - D(S,P)$ because $D(S,P) < D(T,P)$, therefore the gain in the distance should be larger when the new tweet is added to S.
\end{itemize}
Therefore the function D(.) is submodular as it satisfies both properties. We also know that the Entropy function H(.) is submodular and the summation of submodular functions is submodular~\cite{Krau-Golo12}. Therefore, the objective function in Equation~\ref{eq:obj} is submodular. Maximizing a submodular objective function under the cardinality constraint is NP-hard~\cite{Krau-Golo12}. Therefore, event detection in streams where $m$ is unknown and cannot be determined in a timely fashion, is at least as hard. Later, we will describe our approach which approximates the objective function by assigning tweets to the most similar cluster and determines events using the cluster's user diversity.

\section{Identifying Events}
\label{sec:approach}
During a real-world event, people use Twitter to tweet and retweet their experiences. Therefore, the information from these users can be aggregated/clustered to detect events. For streaming Twitter data, however, extra care has to be taken because (1) streaming tweets arrive continuously, traditional multi-pass clustering cannot handle streaming data, and (2) the informal language of tweets defies the standard preprocessing of text corpora such as stemming and vectorization.

To handle high volume and high velocity streaming data, clustering approaches must return the clusters in a single pass. Therefore, we require a clustering approach, which does not require multiple passes over data and which can continuously process the tweets as they arrive. In this paper, we use the single-pass clustering technique described in~\cite{Rijs79}, to group related tweets into clusters as they arrive. This incremental clustering approach continually processes tweets as follows:
\begin{enumerate}
\item A tweet is compared with all the candidate clusters.
\item The tweet is added to the closest cluster, if the distance to the cluster is below a threshold.
\item Otherwise, a new cluster is created and the tweet is added to it as its first member.
\end{enumerate}

To cluster the tweets, we must choose a distance measure appropriate for the characteristics of a tweet stream. We require that the distance measure: be scalable to high-volume streaming data; avoid the need for expensive data transformations, be robust to informal language, and avoid the determination and maintenance of a vocabulary as the language is constantly evolving. Next, we discuss the compression based distance, which addresses these requirements.

\subsection{Tackling Data Informality}
Compression distance computes the distance between two texts by measuring the compression gain obtained on the merging of the two texts. It has been shown to be both efficient and effective for clustering text in~\cite{Keog-etal04}. Additionally, compression based distance has been shown to be effective on multilingual text. Although only discussed in the context of traditional documents, this distance measure is able to handle tweets due to its design. On Twitter, the advantage of compression distance over traditional distance measures such as cosine similarity, is its ability to handle the informal and evolving language in tweets. While cosine similarity requires the maintenance of a vocabulary and data transformation, compression distance can be directly applied to text.

Compression distance is an approximation of the Kolmogorov complexity proposed in~\cite{Keog-etal04}. 
In this paper, we consider each tweet as a document. If $C$ is any compressor, and $C(x)$ is the size of the compressed tweet $x$. Then the distance between two tweets $x$ and $y$, $D(x,y)$ is
\begin{equation}
D(x,y) = \frac{C(xy)}{C(x)+C(y)},
\label{eq:compdist}
\end{equation}
where C(xy) is the compression achieved by merging the two tweets. 

Using the above definition of compression distance between tweets, we can compute the distance between two events $D(e_1,e_2)$ as the maximum pairwise distance between any pair of tweets in the clusters.

\textbf{Choosing the Compressor:} Existing literature recommends choosing a compressor appropriate for the problem domain. In this paper, we compared 3 compression algorithms: DEFLATE, Gzip, and QuickLZ for compression speed and compression ratio using a random set of 20,000 tweets. We found that DEFLATE was the most efficient algorithm in both criteria. Therefore, we will employ it as the compressor \textbf{C} in Equation~\ref{eq:compdist}.

\subsection{Scaling to High Volume Data}
Twitter users currently generate more than 400 million tweets a day~\cite{Tsuk13}. Using publicly available Twitter APIs, one can access a sample of (1\%) tweet stream, which can lead to as many as several million tweets a day. Thus, detecting events in a stream necessitates a scalable solution. Here, we present detailed solutions to scalability.

Events are dynamic and it is essential to consider the temporal evolution of the events in the task of event detection on streaming data. The incorporation of a temporal model into event detection has the following advantages: 
\begin{itemize}
\item capturing evolving events, and 
\item improving the efficiency of event detection. 
\end{itemize}

A cluster representing an event can be considered to be active or inactive at any given time based on the arrival of new tweets. Here, we propose a temporal model which can be used to make this decision. We model events as a Poisson process, which have been traditionally used to model the number of objects in an event at time $t$. In a Poisson process, the rate of arrival of tweets can be modeled as an exponential distribution. This rate is represented by the parameter $\lambda$. Let's consider a random variable $X$, where $X$ measures the time between successive tweets. The variable $X$ is modeled as an exponential distribution with parameter $\lambda$ as
\begin{equation}
X \propto exp(\lambda).\\
\label{eq:hierarch}
\end{equation}

Given an event $e$ and the number of tweets in each time interval in the event $x_1, x_2,..., x_n$, the likelihood function for the inter-arrival time is
\begin{equation}
L(\lambda|x_{1},x_{2}, ...,x_{n}) \propto f(x_1, x_2,..., x_n|\lambda) \propto \prod_{i=0}^{n}{\lambda e^{-x_i \lambda}}.
\end{equation}

To obtain the $\lambda_{MLE}$, we take the derivative of the log-likelihood with respect  to $\lambda$ and set it to zero. Then, $\lambda_{MLE} = \frac{1}{\bar{x}}$, where $\bar{x}$ is the mean of the distribution. For each cluster $c$, if a tweet does not arrive in $\lambda_c$ time units, the current estimate for cluster $c$, then $c$ is considered inactive and removed from memory. The estimate for $\lambda_c$ is updated every time a new tweet is added to the cluster.

\subsection{Identifying Events from Clusters}
Tweets are noisy and not every tweet in the stream is expected to be part of an event. Therefore, not every cluster identified by the algorithm can be an event. The volume of a cluster can help us identify events, but this is susceptible to noise. As a crowdsourced information sharing platform, the diversity of the users (or the number of unique users) who publish tweets in a cluster lends credibility to the information within the cluster. Therefore, we measure the user diversity of a cluster to determine whether it is an event. A cluster is classified as an event, if its Diversity Score $H(c)$ is above the \textit{Diversity Threshold} $H_{t}$. 

\begin{algorithm}[t]
\caption{Event Detection in Twitter Streams}
\KwIn{A stream of tweets $T$ and the Cluster Limit ($k$), the Tweet Limit ($l$), the Distance Threshold ($D_t$), and the Diversity Threshold ($H_t$).}
\KwOut{Detected events $E$.}
$E \gets \{\}$\; 
$C \gets \{\}$\;
\While{tweet $t \in T$}{
	Identify active cluster $c \in C$, where $D(t,c) \leq D_t$\;
	\If{c exists}{
		Add $t$ to $c$\;
		Update expected time of next tweet $\lambda_c$\;
		Update User Diversity ($H(c)$) of cluster $c$\;
		\If{$H(c) \geq H_t$}{
			Mark cluster as an event, Add $c$ to $E$\;
		}
	}
	\Else {
		Create new cluster $c$ with $t$ as its first member\;
		Add $c$ to $C$\;		
	}
}
\label{alg:event}
\end{algorithm}

\section{Event Detection Framework}
Using the strategies to handle informal language in tweets, temporal dynamics of events, and handling noise, we can detect events in Twitter streams using Algorithm~\ref{alg:event}. To improve the efficiency of the algorithm and to scale it to large Twitter streams we propose two heuristics: \\
\textit{Cluster Limit: }The assignment of tweets to clusters requires a comparison with currently active clusters, but sequential search of all active clusters can be prohibitive. As a tweet is more likely to be similar to clusters with overlapping content we limit the comparisons to these candidate clusters. These clusters are identified by aggregating, sorting, and ranking clusters according to their overlap with the tweet. Then, we pick the top $k$ clusters as the candidate clusters. $k = 100$, was empirically found to be effective in discovering reasonable clusters without sacrificing the speed of the algorithm.\\
\textit{Tweet Limit: }The distance of a tweet to an event is computed as the maximum pairwise distance with the tweets contained in the event. Due to the timely nature of tweets, we propose to restrict the comparisons ordered by recency. This could be effective when clusters represent events which span an extended period of time and contain a large numbers of tweets. We propose to restrict the comparisons to at most $l$ recent tweets in a cluster. In our implementation, we set $l=1000$.
\textbf{Time Complexity:} Given the number of tweets in the stream as $N$, the Cluster Limit ($k$) and the Tweet Limit ($l$), the time complexity of our algorithm is $O(Nkl)$. The most expensive operation in our algorithm is the assignment of tweets to clusters. For every tweet in the stream, it needs to be compared to at most $l$ tweets in $k$ clusters. As the values of $k$ and $l$ are much smaller than $N$, the algorithm allows us to process the tweets in near real-time. In the later sections, we will present empirical evidence of the algorithm's efficiency.

\textbf{Selection of Parameters:} Two thresholds are used in our framework to identify events. First, the distance threshold $D_t$ is used to determine assignment of tweets to clusters. In a study on 20,000 random tweets, we found that the average self-similarity of tweets was 0.54 and a value of 0.8 was empirically found to be a suitable value to obtain reasonable clusters. Second, the diversity threshold $H_t$ is used to decide which clusters can be labeled as events, as noise is a problem in tweet streams. Volume or the number of tweets in a cluster is also an important factor in determining whether a cluster is an event.  Ideally, we would like clusters to contain many tweets and have high user diversity. This threshold was set empirically as outlined later.

In the next section, we present evaluation results along: 1) scalability to high volume and high velocity streams, and 2) quality of the detected events. 

\section{Evaluation Strategy}
\label{sec:evalstrat}
There are two specific challenges in evaluating events from Twitter: 1) Unlike traditional media such as broadcast news, where every event is reported, on Twitter there is less likelihood of finding tweets related to minor events, and 2) While traditional research on event detection has relied upon the availability of labeled corpora such as the TDT corpus for evaluation, no such corpus exists for Twitter. Due to the lack of ground truth the exact number or nature of the events is not easily available and manual labeling of a large Twitter dataset is expensive. Twitter streams can be collected in two forms:
\textbf{topic streams} containing tweets related to a specific topic, where the number and type of events can be verified using external sources, and \textbf{random streams}, which contain randomly sampled tweets, where the number and type of events must be manually determined. In this section, we evaluate the proposed approach on both types of streams.

\subsection{Detecting Events in Topic Streams}
As a representative topic stream, we introduce the Earthquake topic stream which consists of tweets related to earthquakes around the world.
\\
\textbf{Earthquakes:} due to the existing research demonstrating the use of Twitter during earthquakes~\cite{Saka-etal10}, \cite{Mend-etal10}, we collected tweets referring to earthquakes between June, 2011 to May, 2012 by monitoring the hashtags: \#earthquake, \#terremoto, and \#quake. The data comprises of 1,007,417 tweets from 317,564 users. 

To identify the real-world events spanned in this dataset, we must find an independent and external source, which can provide the ground truth at a reasonable cost as manual annotation is not practical. Towards this, we selected the major earthquakes in 2011~\cite{Wiki11b} and 2012~\cite{Wiki12c} on Wikipedia as the ground truth on the nature of the events in the dataset. These reports were manually compiled from several major news sources. In this paper, we focus our effort on the days when a major earthquake resulted in at least 10 casualties, which are summarized in Table~\ref{tab:eqgt}. For most events in 2011, only a few hundred tweets were collected which might be due to the popularity of regional hashtags. Therefore, we set $H_{t} = 5$ for this dataset.\\

\begin{table}[t]
\caption{Major earthquakes in 2011 and 2012}{
\scriptsize
\begin{tabular}{|p{2cm}|p{2.5cm}|p{1.2cm}|p{1cm}|}
\hline
\textbf{Day(UTC)}&\textbf{Location}&\textbf{Magnitude}&\textbf{Death Toll}\\
\hline
Jul 19, 2011&Fergana Valley&6.2&14\\
\hline
Sept 5, 2011&Aceh, Indonesia&6.7&12\\
\hline
Sept 18, 2011&India-Nepal border&6.9&111\\
\hline
Oct 23, 2011&Van, Turkey&7.1&684\\
\hline
Nov 9, 2011& Van, Turkey&5.7&40\\
\hline
Feb 6, 2012&Visayas, Philippines&6.7&113\\
\hline
Apr 11, 2012&Aceh, Indonesia&8.6&10\\
\hline
May 20, 2011&Emilia-Romagna, Italy&6.1 \& 5.8&27\\
\hline
\hline
\end{tabular}
}
\label{tab:eqgt}
\vspace{-.5cm}
\end{table}

\begin{table}[t]
\caption{Efficiency of event detection: Earthquake}{
\scriptsize
\begin{tabular}{|p{1.5cm}|p{1cm}|p{1cm}|p{1.3cm}|p{1.5cm}|}
\hline
\textbf{Day}&\textbf{\#tweets}&\textbf{Total processing time (Min)}&\textbf{Collection rate (Tweets/Min)}&\textbf{Processing rate (Tweets/Min)}\\
\hline
Jul 19, 2011&880&0.04&0.613&23,498.00\\
\hline
Sept 5, 2011&2,712&0.18&1.88&14,788.69\\
\hline
Sept 18, 2011&465&0.02&0.32&18,699.73\\
\hline
Oct 23, 2011&5,253&0.49&3.65&10,646.89\\
\hline
Nov 9, 2011&2,712&0.17&1.89&15,611.63\\
\hline
Feb 6, 2012&13,586&4.79&13.72&2,834.92\\
\hline
Apr 11, 2012&28,182&10.61&19.57&2656.06\\
\hline
May 20, 2012&20,509&6.40&14.33&3,204.44\\
\hline
\hline
\end{tabular}
}
\label{tab:eqeff}
\vspace{-.5cm}
\end{table}

\subsubsection{Evaluating Scalability}
To verify that our approach is scalable, we evaluated the rate at which the tweets in our dataset were generated and the time required by the proposed framework to identify events. Table~\ref{tab:eqeff} compares the measurements for the Earthquake dataset. Column 4 describes the rate at which tweets were collected and Column 5 describes the rate at which tweets were processed. We find that the proposed approach is capable of handling high volume topic-specific Twitter streams by being able to process the tweets at a rate which is significantly higher than the rate at which tweets were generated.

\begin{table}[t]
\centering
\caption{Events detected in the Earthquake dataset}{
\scriptsize
\begin{tabular}{|p{1.35cm}|p{1.6cm}|p{4.2cm}|}
\hline
\textbf{Day}&\textbf{Earthquake Location}&\textbf{Key Event Terms}\\
\hline
Sept 5, 2011&Indonesia&sumatra, western, indonesian, island, \#breakingnews\\
\hline
Oct 23, 2011&Turkey&\#turkey, eastern, turkey, magnitude, news\\
\hline
Nov 9, 2011&Turkey&turkey, eastern, magnitude, rocks, usgs\\
\hline
Feb 6, 2012&Philippines&pray, visayas, philippines, struck, earlier\\
\hline
Apr 11, 2012&Indonesia&\#indonesia, tsunami, magnitud, indonesia, sacudió\\
\hline
May 20, 2012&Italy&sentito, emilia, sono, cosa, chies\\
\hline
\hline
\end{tabular}
}
\label{tab:eqevents}
\end{table}

\subsubsection{Quality of Detected Events}
Detected events are typically described using the frequent keywords from event tweets~\cite{Yang-etal98,Petr-etal10,Fung-etal05}. Therefore, we extracted the top keywords of each event as its description to verify whether they matched the ground truth. In Table~\ref{tab:eqevents}, we present the most representative event corresponding to the events in Table~\ref{tab:eqgt}. We also observed that the proposed approach was able to discover the evolution of events, which are represented by sub-events, which we will revisit later.


To quantify the effectiveness of our approach in detecting events, we compute the $F_1$ score which captures both the Precision and Recall. Precision is computed as the number of detected events that match the ground truth including sub-events. Recall is computed as the number of events from the ground truth which were successfully detected. 
The $F_1$ Score for the Earthquake dataset was 0.77. 

\subsection{Detecting Events in a Random Stream}
Twitter streams can also be collected without any topic bias using the Twitter Sample API\footnote{https://dev.twitter.com/docs/api/1.1/get/statuses/sample}. Using this API, we can retrieve a 1\% random sample of the Twitter stream. The tweets in such streams include interpersonal conversations and discussions of real-world events. The task of event detection is harder in this case due to the presence of noise. To verify that our approach can be successfully applied to random streams, we collected sampled tweets from 11:02 AM on Apr 15 to 9:16 AM on Apr 16, 2013. The data consisted of 4,212,333 tweets from 3,322,379 users. As there is no ground truth for these days, we will test the effectiveness of our framework by verifying that we can detect the top stories of the day. We begin by establishing the scalability of our approach. Here we set $H_t=6.3$ due to the larger volume.

\subsubsection{Evaluating Scalability}
\begin{figure}[t]
\centering
\includegraphics[scale=0.32]{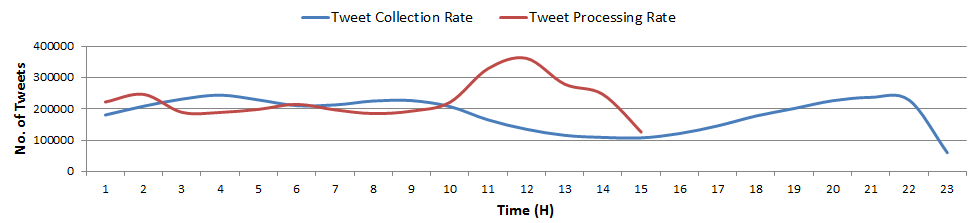}
\caption{A comparison of the tweet collection rate and the tweet processing rate in the Random dataset}
\label{fig:sampleprocessing}
\end{figure}

As in the case of the topic specific dataset, we test the efficiency of the proposed approach on a random stream by comparing the tweet generation rate and tweet processing rate. A comparison of these measurements is presented in Figure~\ref{fig:sampleprocessing}. The figure clearly shows that the processing speed for a majority of the collection period was on par with the collection speed and it often exceeded the tweet collection rate significantly. Nevertheless, the proposed approach was able to detect events in near-real time. This shows that our approach can be efficiently applied to a random Twitter stream.

\subsubsection{Quality of Detected Events}
As manually labeling the tweets is not practical, we evaluate the quality of the events based on the coverage of the two major events which occurred during this time period. From the random stream, we detected 167 events. A manual investigation of the events revealed 4 major kinds of events: events related to the Boston bombing incident, events related to the Presidential elections in Venezuela, events discussing a music festival, and finally events which represented banal Twitter chatter. An example of events expressing banal Twitter chatter included tweets from the fans of Justin Bieber, which resembled the characteristics of an event, but did not refer to a specific event. In Table~\ref{tab:randomevents}, we present examples of the two main types of events: Boston marathon bombing and the Venezuelan Presidential elections. The first event discusses the controversy surrounding the counting of votes in the Presidential elections in Venezuela held on Apr 14, 2013~\cite{Wiki13}. The other two events are related to the Boston marathon bombing incident. The first event contains reports of the bombing itself and the second event references the reactions of the Twitter users. Our results show that we are able to detect reasonable events in the presence of large amount of noise.

\begin{table}[t]
\centering
\caption{Events detected in the Random dataset}{
\scriptsize
\begin{tabular}{|p{3.5cm}|p{3.5cm}|}
\hline
\textbf{Event}&\textbf{Top Keywords}\\
\hline
Venezuelan Presidential elections voting controversy&votos, capriles, esto, \#caprilesganótibisaymintió, fraude\\
\hline
Boston marathon bombing incident&marathon, boston, explosion, finish, line\\
\hline
Support for the bomb victims starts pouring in&boston, marathon, explosion, heart, bombing\\
\hline
\hline
\end{tabular}
}
\vspace{-.5cm}
\label{tab:randomevents}
\end{table}

\section{Discussion}
While few approaches exist to capture events in a streaming environment, the \textit{Threading} technique proposed in~\cite{Petr-etal10} is the closest. Using the configuration recommended by the authors, we applied this technique to the Earthquake dataset. First, we compare the scalability of the two approaches. The results for this experiment are presented in Table~\ref{tab:peteff}. A comparison against our approach in Table~\ref{tab:eqeff} shows that our approach can process and detect events faster. Next, we evaluate the quality of the events. On all days, the Threading approach detected a greater number of events. Even using the ranking strategy proposed by the authors to retrieve the top 10 fastest growing events, we found that the $F_1$ score for the Threading technique was 0.64 compared to 0.77 for the proposed approach. As the Earthquake dataset was the smallest among our datasets, the results show that our framework outperforms this approach. The proposed approach also successfully removed noisy tweets. 
\begin{table}[t]
\caption{Efficiency of Threading technique: Earthquake}{
\scriptsize
\begin{tabular}{|p{1.4cm}|p{1cm}|p{1.2cm}|p{1.2cm}|p{1.2cm}|}
\hline
\textbf{Day}&\textbf{\#tweets}&\textbf{Total Processing Time (Min)}&\textbf{Tweet collection rate (Tweets/Min)}&\textbf{Processing rate (Tweets/Min)}\\
\hline
Jul 19, 2011&880&1.11&0.613&793.40\\
\hline
Sept 5, 2011&2,712&3.99&1.88&678.68\\
\hline
Sept 18, 2011&465&0.88&0.32&527.10\\
\hline
Oct 23, 2011&5,253&2.65&3.65&1,984.97\\
\hline
Nov 9, 2011&2,712&2.54&1.89&1,068.13\\
\hline
Feb 6, 2012&13,586&38.36&13.72&354.19\\
\hline
Apr 11, 2012&28,182&135.27&19.57&208.34\\
\hline
May 20, 2012&20,509&210.32&14.33&97.51\\
\hline
\hline
\end{tabular}
}
\label{tab:peteff}
\vspace{-.1in}
\end{table}

An additional advantage of the proposed framework is that it can detect the evolution of events in dynamic Twitter streams. The inclusion of a \textit{temporal model} allows us to identify sub-events within a larger event. For example, we can not only detect that an earthquake has occurred, but also detect the topics that emerge as a result of the earthquake, such as damage reports as seen in Table~\ref{tab:eqevo}, where we present 5 events from the tweets generated during the Indonesian earthquake on April 11, 2012.

\begin{table}[t]
\centering
\caption{Evolution of events on April 11, 2012}{
\scriptsize
\begin{tabular}{|p{3.8cm}|p{3.8cm}|}
\hline
\textbf{Event}&\textbf{Top Keywords}\\
\hline
Earthquake strikes Indonesia. Tsunami alert is issued&tsunami, \#indonesia, \#sumatra, scossa, allarme\\
\hline
Tremors felt in India&felt, singapore, thailand, indonesia, \#tremors\\
\hline
Tsunami alert in Indian Ocean&tsunami, indian, ocean, move, alert\\
\hline
Sea water receding near the epicenter&aceh, quake, water, receding, island\\
\hline
Reports emerge that a tsunami is less likely&\#indonesia, tsunami, moved, horizontally, vertically\\
\hline
\hline
\end{tabular}
}
\label{tab:eqevo}
\vspace{-.1in}
\end{table}

\section{Related Work}
\label{sec:related}
Event detection in traditional media is also known as Topic Detection and Tracking (TDT) and a pilot study on this task is presented in~\cite{Alla-etal98a}. In~\cite{Yang-etal98}, news articles were modeled as documents to detect topics. The documents were transformed into vector space using the TF-IDF and two clustering approaches were evaluated: Group-Average Agglomerative Clustering (GAAC) for retrospective event detection, and Incremental Clustering for new event detection. The authors concluded that the task of new event detection was harder. In~\cite{Alla-etal98b}, the authors focused on online event detection. The authors approached the problem as a document-query matching problem. A query was constructed using the $k$ most frequent words in a story. If a new document did not trigger existing queries then it was considered to be part of a new event. In~\cite{Fung-etal05}, the authors addressed the problem of detecting \textit{hot} bursty events. They introduced a new parameter-free clustering approach called feature-pivot clustering, which attempted to detect and cluster bursty features to detect hot stories. 

An attempt to detect earthquakes using Twitter users as social sensors was carried out by in ~\cite{Saka-etal10}. The temporal aspect of an event was modeled as an exponential distribution, and the probability of the event was determined based on the likelihood of each sensor being incorrect.~\cite{Beck-etal10} tackled event detection in Flickr. The authors leveraged the meta data of images to create both textual and non-textual features and proposed the use of individual distance measures for each feature. These features were used to create independent partitions of the data and finally the partitions were combined using a weighted ensemble scheme to detect event clusters. In~\cite{Weng-Lee11}, the authors constructed word signals using the Wavelet Transformation and used a modularity-based graph partitioning approach on the correlation matrix to get event clusters.~\cite{Li-etal12} identified bursty segments in tweets and clustered the segments to identify events.

Few existing approaches are designed for streaming Twitter data and even fewer are scalable to real-time streams. In~\cite{Sayy-etal09}, the authors converted a stream of blog posts into a keyword graph, where nodes represented words and links represented co-occurrence. Community detection methods were applied on the graph to detect communities of related words or events. In~\cite{Zhao-etal07}, the authors model the social text streams including blogs and emails as a multi-graph and cluster the streams using textual, temporal, and social information to detect events. A hybrid network and content based clustering approach was employed  in~\cite{Agga-Subb12} to identify a fixed number of events in a labeled Twitter stream containing tweets from two events. Generally, the number of events is not known beforehand and obtaining the user network adds significant overhead, thus adding to the complexity of this method. In~\cite{Petr-etal10}, the authors recognized the need for faster approaches for first story detection in streams. The authors proposed a two-step process to identify first stories in streaming data. First, the nearest neighbor of each tweet is identified using locally sensitive hashing in constant time and space. Second, a clustering approach called Threading is applied to group related tweets into event clusters. The first tweet in a thread is presented as the first story and the thread itself is considered an event. 

\section{Conclusions and Future Work}
In this work, we presented a novel approach to detect events in informal and high volume Twitter streams. The results demonstrate that the proposed approach can handle the informality of language in Twitter streams, through the use of compression distance.  We found that the proposed approach is capable of handling dynamic streams, where the number of events is unknown or cannot be practically determined in near real-time. The proposed User Diversity measure is also able to successfully filter noise in Twitter streams. Through experiments we demonstrated that the proposed approach is efficient and is able to capture reasonable events in topic streams and random streams on Twitter.

Event detection has several potential applications, which we intend to investigate as part of our future work. Investigating the relationship between an event's rate of growth and its impact in the real-world is one. Another direction of future study is the categorization of events based on two characteristics: the volume of the event defined by the number of tweets, and the rate of the event. By organizing the events in this fashion, we can provide a value added service to a user by facilitating the tracking of specific types of events.

\section*{Acknowledgments}
This work was sponsored, in part, by the Office of Naval Research grant N000141410095.

\bibliographystyle{aaai}
\bibliography{icwsm}
\end{document}